# Screening limited switching performance of multilayer 2D semiconductor FETs: the case for SnS

Sukrit Sucharitakul,[a] U. Rajesh Kumar,[bc] Raman Sankar,[de] Fang-Cheng Chou,[d] Yit-Tsong Chen,[bc] Chuhan Wang,[f] Cai He,[f] Rui He,[f] and Xuan P. A. Gao[*a]

Gate tunable p-type multilayer tin mono-sulfide (SnS) field-effect transistor (FET) devices with SnS thickness between 50 and 100 nm were fabricated and studied to understand their performances. The devices showed anisotropic inplane conductance and room temperature field effect mobilities ~5 - 10 cm$^2$/Vs. However, the devices showed an ON-OFF ratio ~10 at room temperature due to appreciable OFF state conductance. The weak gate tuning behavior and finite OFF state conductance in the depletion regime of SnS devices are explained by the finite carrier screening length effect which causes the existence of a conductive surface layer from defects induced holes in SnS. Through etching and n-type surface doping by $Cs_2CO_3$ to reduce/compensate the not-gatable holes near SnS flake's top surface, the devices gained an order of magnitude improvement in the ON-OFF ratio and hole Hall mobility ~ 100 cm$^2$/Vs at room temperature is observed. This work suggests that in order to obtain effective switching and low OFF state power consumption, two-dimensional (2D) semiconductor based depletion mode FETs should limit their thickness to within the Debye screening length of carriers in the semiconductor.

## Introduction

Since the work by Novoselov and coworkers on exfoliable 2D van der Waals (vdW) materials[1] and studies that demonstrated Dirac electrons in graphene,[1-2] scientists and engineers have been conducting enormous amount of research on the topic. A wide variety of 2D materials were explored and studied for their exotic electrical and optical properties, in particular, transitional metal dichalcogenides (TMDs)[3-7] such as $MoS_2$[8-10], $MoSe_2$[11,12], $WS_2$[13,14] and $WSe_2$[15,16]. In addition to TMDs, the pursuit of 2D semiconductors for high performance electronics or optoelectronics is extended to non-TMDs such as phosphorene[17,18] and III-VI materials (e.g. InSe).[19,20] Among the numerous 2D semiconductors that have been explored, many high quality n-type materials or device structures have been developed. Achieving high performance p-type devices has been more challenging. Notable recent progresses in this area are black phosphorus devices[17,18] and novel p-type 2D contact scheme for hole injection into TMD devices.[21]

Tin monosulfide (SnS), a IV-VI compound 2D semiconductor, has a layered Pnma crystal structure. In the past, due to its 1.07 eV band gap and its high optical absorption coefficient above band gap,[22,23] SnS has been mainly studied for solar cell applications.[23] More recently, the electronic and thermal properties of bulk SnS and SnSe have attracted increasing attention for thermoelectric applications.[24-26] So far, transport study on nanoscale SnS and SnSe crystals is very limited.[27] Given that bulk SnS has p-type nature (due to tin vacancies) and hole mobility of 90 cm$^2$/Vs,[28] SnS appears to be a good candidate for use in p-type 2D nanoelectronic devices. The anisotropic crystal structure of SnS also suggests that the transport properties of SnS could be strongly anisotropic within the 2D plane, similar to black phosphorus.[26] However, there has been no carrier transport or field effect device studies on nanoscale SnS. In this work, nanoflakes were exfoliated from bulk SnS single crystal and p-type SnS FET devices were successfully fabricated and studied. We found that due to the high intrinsic p-type doping and strong carrier screening in as grown SnS, the field effect of gate is not sufficient to tune the carrier transport throughout the whole thickness (ca. 50-100 nm) of the devices and renders poor switching behavior. By means of surface doping to compensate the intrinsic p-type holes or etching to reduce the sample thickness, devices with hole Hall mobility ~ 100 cm$^2$/Vs (at room temperature) and improved ON-OFF ratio (~ 100 at room temperature) were realized. In addition to gaining a general insight on the role of carrier screening as the limiting factor in the switching performance of multilayer 2D semiconductor depletion mode FETs, this work paves a way towards high performance 2D electronic devices based on IV-VI semiconductors and the theoretically envisioned novel valleytronics based on single layer SnS.[29]

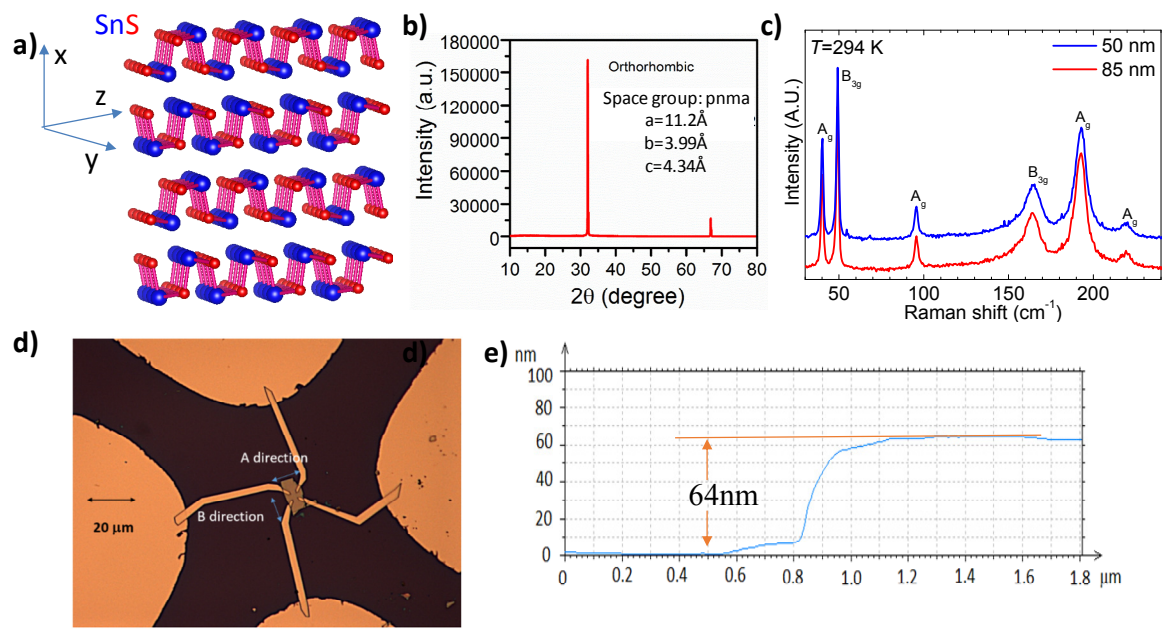

**Fig. 1.** a) Crystal structure of SnS (Sn in blue and S in red). b) XRD results of a bulk SnS single crystal used. c) Raman spectra from 50 and 85 nm SnS flakes. d) Optical image of a SnS four-probe device. e) AFM height profile of an exfoliated SnS flake used to fabricate device.

## Results and discussion

The layered vdW crystal structure of SnS is illustrated in the scheme shown in Fig. 1a. The structure is highly anisotropic, both along the direction perpendicular to the 2D layer as well as within the two-atom thick 2D layer. Within the 2D layer, the atoms are arranged in the armchair fashion along the z (or c) direction (as shown in Fig. 1a) while the y (or b) direction is the zigzag direction. The x-ray diffraction (XRD) data of the bulk SnS crystal used in this work are shown in Fig. 1b, confirming its single crystalline quality. Fig. 1c shows Raman spectra from 50 nm and 85 nm SnS flakes. Raman spectrum from each sample shows six well-defined peaks, among which four are $A_g$ modes and two are $B_{3g}$ modes. The frequencies of these modes are in excellent agreement with those reported for bulk SnS,[30, 31] confirming that our samples are in orthorombic phase. When exfoliated into small multilayer flakes, SnS samples often appear in a shape near a rectangle. To identify the anisotropic electronic properties of SnS, we distinguish the short and long directions of SnS nanoflake samples as the 'A' and 'B' directions, respectively, as noted in the picture of a representative device in Fig. 1d. The shorter flake length and higher mobility measured along the 'A'-direction (to be discussed later) suggest that the 'A'-direction corresponds to principal axis b while B corresponds to principal axis c.[26] Electron-beam lithography was used to fabricate four-probe devices in the van der Pauw configuration (Fig. 1d). Atomic force microscope (AFM) was used to identify the thickness of SnS flakes studied. An example of the height profile of a 64 nm thick SnS flake imaged by AFM is given in Fig. 1e.

The sample was loaded into a Physical Properties Measurement System (PPMS) for electrical measurements. As demonstrated by the linear relation between the source-drain current $I_{sd}$ and the source-drain voltage $V_{sd}$ in Fig. 2a, Ohmic contacts were obtained between Ni source/drain contacts and SnS nanoflake. Moreover, the four-probe van der Pauw geometry allowed comparing SnS nanoflake's intrinsic conductance along two orthogonal crystal directions without the influence of contact resistance. Fig. 2b presents the four-probe conductance of a SnS nanoflake measured along the A-direction or B-direction as a function of backgate voltage $V_g$ at $T=300$ K. The decreasing trend of conductance vs. $V_g$ indicates the p-type nature of the SnS flake. The conductance along the A-direction is generally found to be higher than that measured along B-direction in our van der Pauw devices. For the device in Fig. 2b, the conductance ratio between these two directions is about 2.5 (Fig. 2b inset). This is consistent with the anisotropic lattice structure and electronic structure in the 2D plane of SnS.[26] It is noteworthy from Fig. 2b that the device could not be completely turned off at high positive gate voltages (i.e. there is a non-negligible conductance in the OFF state of the device at high $V_g$). Fig. 2c further shows the four-probe conductance $G$ vs. $V_g$ along the B direction at different temperatures. The $G(V_g)$ data at different $T$ show that raising temperature increases the overall conductance of the device and the device tends to saturate towards a lower OFF state conductance at lower $T$ as indicated by the dashed horizontal lines in Fig. 2c and d. The strong influence of temperature on devices' ON-OFF ratio can be more clearly seen in Fig. 2d which displays the $G(V_g)$ data in semi-log scale. At room temperature, the ON-OFF ratio is ~10 and lowering $T$ to 250 K enhances the ON-OFF ratio to ~50. Fig. 2c and d also show noticeable hysteresis effect in the $G$ *vs.* $V_g$ sweep loop, indicating the existence of charge trap states.[9,32]

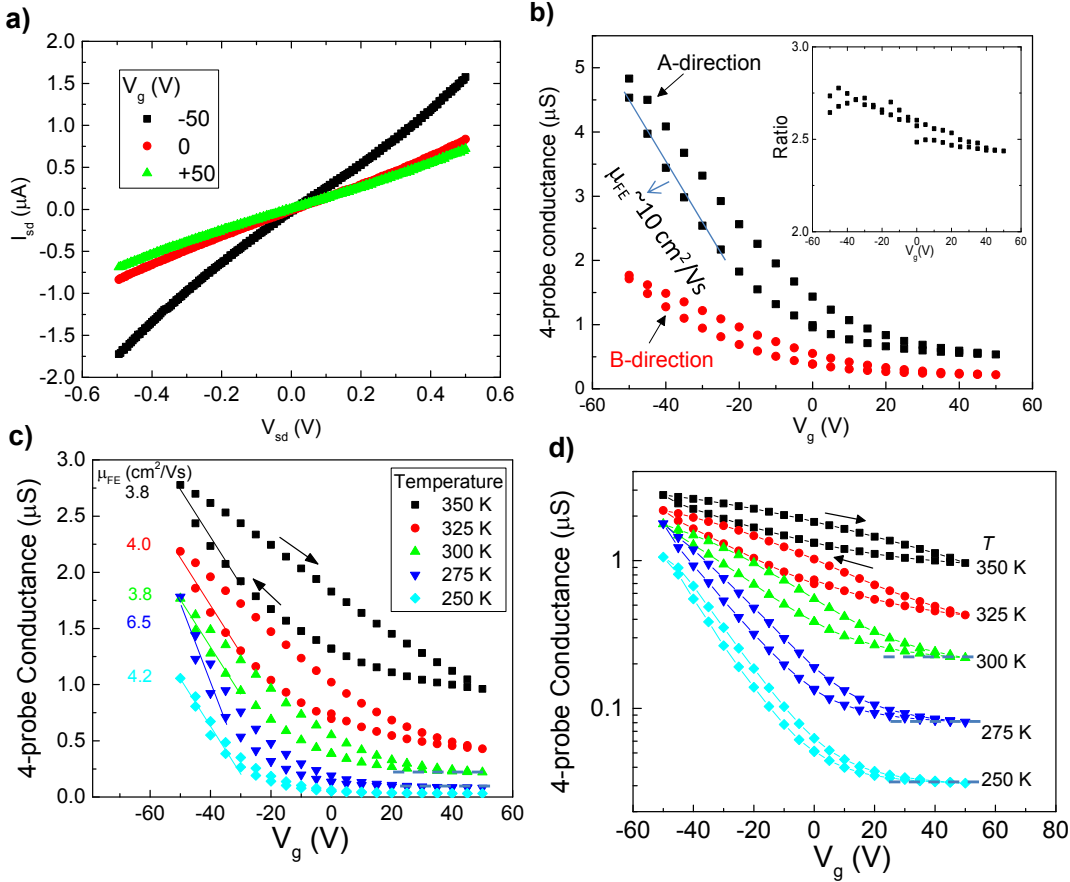

**Fig. 2.** a) $I_{sd}$-$V_{sd}$ plot of a 60 nm thick multilayer SnS device at 300 K. b) Four-probe conductance *vs.* $V_g$ along two perpendicular directions in the 2D plane at 300 K (inset: conductance ratio between the two directions). c) Four-probe conductance along the B direction *vs.* $V_g$ at different temperatures and d) Data in c) shown in semi-log scale.

From the quasi-linear regime of the $G(V_g)$ plots at negative $V_g$, the field effective mobility $\mu_{FE}$ can be extracted using the relation $\mu_{FE} = \frac{g_m}{C_g}/(\frac{\pi}{\ln(2)})$, where $g_m$ is the trans-conductance or the slope of conductance *vs.* $V_g$ while $C_g$ is the gate capacitance and $\frac{\pi}{\ln(2)}$ is the correction factor in the van der Pauw method. $\mu_{FE}$ can be estimated to be around 4-7 cm$^2$/Vs along the B direction with no obvious $T$-dependence in the temperature range 250 - 350 K. The $\mu_{FE}$ along the higher mobility A direction is around 10 cm$^2$/Vs. Compared to the ~90 cm$^2$/Vs hole mobility in bulk SnS, this $\mu_{FE}$ is almost an order of magnitude smaller. Our gate dependent carrier density measurement from the Hall effect (Fig. 3a) showed that the gate capacitance is significantly over-estimated in the parallel plate capacitance model, due to the disregard of trap states capacitance. Thus this apparent $\mu_{FE}$ ~4-10 cm$^2$/Vs significantly under-estimates the hole mobility in multilayer SnS. (The extracted Hall mobility for the device in Fig. 2&3 is ~ 30 cm$^2$/Vs along the high mobility direction and ~ 10 cm$^2$/Vs along the low mobility direction. Electronic Supplementary Information Fig. S1.)

The intriguing behavior of gate being able to enhance the carriers but incapable of depleting carriers in SnS was observed in many devices with thickness in the range of 50 to 100 nm. This characteristics was also quite common in other semiconductor nanoscale FET devices with similar thickness, (e.g. InAs nanowire FETs [33]) and poses a limitation on the performance of semiconductor nanoelectronics. In the following, we will show that this phenomenon is originated from the simple fact that carrier screening length $L_D$ in semiconductors is finite such that semiconductor FETs with body thickness larger than $L_D$ has a surface 'dead layer' whose carrier conduction cannot be tuned by the gate.

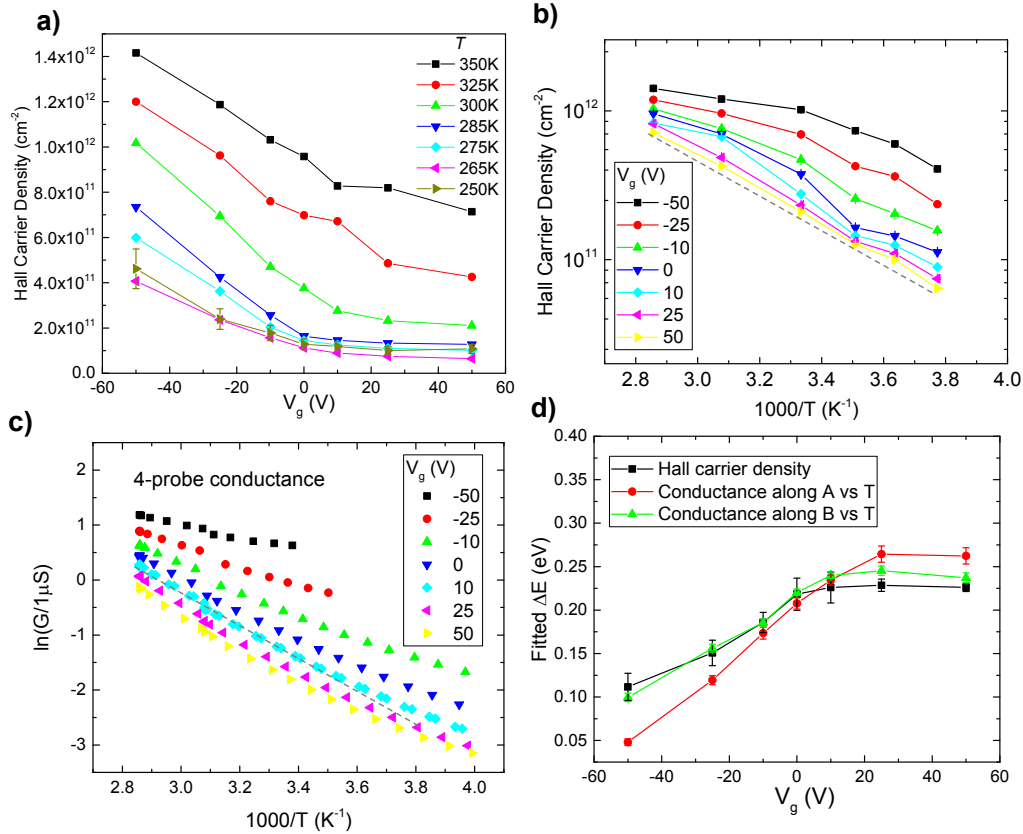

**Fig. 3.** a) Hall carrier density at different applied backgate voltages. b) Arrhenius plot of Hall carrier density for holes *vs.* 1/T. c) Arrhenius plot of the four-probe conductance *vs.* 1/T. d) Fitted activation energy from the temperature dependence of the hole density extracted from the Hall effect or the conductance along the 'A-direction' or 'B-direction' of SnS flake, plotted against the backgate voltage.

To comprehend the origin of carriers and the underlying gate tuning mechanism of the transport properties of the device, Hall measurements at different temperatures and back-gate voltages were conducted. By measuring the Hall resistance $R_{xy}$ at various magnetic field $B$, we observed a linear relation in $R_{xy}$ *vs.* $B$ and the Hall coefficient can be calculated from $R_H = \frac{R_{xy}}{B}$. The hole carrier density can be extracted through relation $p = \frac{1}{eR_H}$ with $e$ being the fundamental electron charge. From the hole carrier density $p$ data shown in Fig. 3a, we see that $p$ increases linearly as $V_g$ reduces towards the negative direction but it saturates towards a residual value at high positive $V_g$, in a similar way to the $G(V_g)$ data in Fig. 2. This suggests that there is a limit on the number of holes that can be removed by applying positive $V_g$ in these SnS flakes. Moreover, for a given $V_g$, there are more carriers at higher temperature, suggesting the thermally excited nature of carriers. Fig. 3b shows the temperature dependence of hole density in the Arrhenius plot. The plot highlights that available hole carrier concentration for electrical transport follows $p(T) \propto \exp(-\frac{\Delta E}{k_B T})$. Similar thermally activated behavior is seen in the Arrhenius plot for the temperature dependent conductance in Fig. 3c. The activation energy $\Delta E$ extracted from fitting the Arrhenius plots of $p(T)$ and $G(T)$ along both the A and B-directions of the SnS flake are displayed in Fig. 3d. We see a good agreement between $\Delta E$ fitted from the thermal activation model for the carrier density and conductance data: the activation energy increases with the gate voltage in the negative $V_g$ regime and stays relatively constant in the positive $V_g$ regime. This trend is again correlated with the gate dependent conductance and hole density in Fig. 2 and Fig. 3a.

The gate tuned thermally activated transport and the behavior of residual conductance in the high $V_g$ regime in these SnS devices can be naturally explained by a simple model in which the holes originate from acceptors with ionization energy $E_a$ ~ 0.22 eV (Fig. 4a) and the carrier screening effect in semiconductor with thickness exceeding the carrier screening length (Fig. 4b). Our estimate of acceptor ionization energy ~ 0.22 eV is based on the activation energy for the carrier density and conductance at $V_g=0$ in Fig. 3d. In the negative gate bias regime, more holes are induced into the bottom surface of SnS flake. This accumulation of holes enhances the conductivity of a layer near the bottom of the flake and the measured conductance is dominated by this bottom layer in which the Fermi energy $E_F$ is shifted closer to the valence band. Thus a reduced activation energy and increased conductance were observed. On the other hand, when a positive $V_g$ is applied to deplete the carriers, holes near the bottom surface are first depleted and the conductance of the sample gradually becomes dominated by the holes near top surface where $p$ and $E_F$ remain constant. Therefore, the activation energy $\Delta E$ appears to be pinned at ~0.22eV by the carriers in the top surface layer.

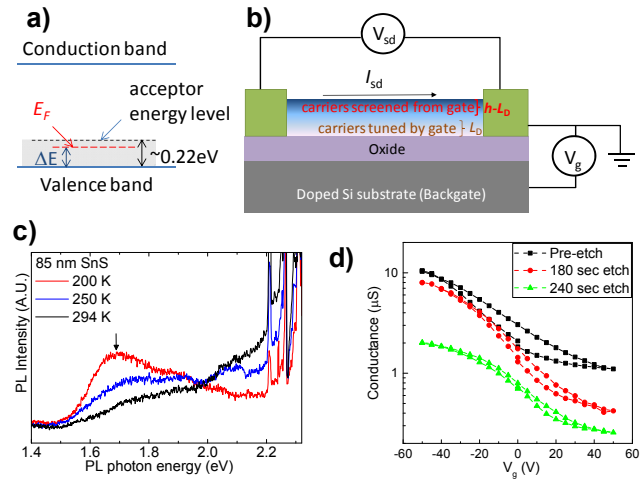

**Fig. 4.** a) Schematic of relative positions of different energy levels. b) Schematic of carrier gating in multilayer SnS FET devices with thickness larger than the screening length $L_D$. Gate is only able to tune the bottom layer of the sample with the rest of sample unaffected by gate voltage. c) PL spectra at different temperatures of an 85 nm thick SnS flake exfoliated onto Si/Si$_3$N$_3$ substrate. The excitation photon energy is 2.33 eV. The superimposed sharp peaks above 2.2 eV are Raman lines from the sample or the substrate. d) Four-probe conductance vs. $V_g$ of a SnS device upon plasma etching to thin the SnS flake (initial thickness was 45nm, final thickness after 240 seconds etching was approximately 20-25nm).

We next discuss the OFF state performance of depletion mode FET devices made from 2D semiconductors with appreciable thickness and doping, using the multilayer SnS devices in this work as an example. With small $V_g$, the depth of bottom surface tuned by gate is approximately the Debye screening length $L_D = \sqrt{\varepsilon k_B T / e^2 p}$ with $\varepsilon$, $k_B$ as the dielectric constant of semiconductor and the Boltzmann constant (Fig. 4b). In very high $V_g$ towards depletion, one reaches the maximum depletion depth when $E_F$ of bottom surface is tuned to the middle of the bandgap at which point the depletion thickness $W_D \approx \sqrt{2\varepsilon k_B T \ln(\frac{N_a}{p_i})/e^2 N_a}$ with $p_i$ and $N_a$ as the intrinsic carrier concentration and acceptor density.[34] Based on the hole density at $V_g$=0 (Fig. 3a) and flake thickness, we estimate the ionized acceptor density $N_a$ to be about $10^{17}$/cm$^3$ in our SnS at room $T$. This leads to a maximum depletion width $W_D$ to be a few tens of nm at 300K.[34] Therefore, when a large gate voltage is used to deplete carriers in doped 2D semiconductors with thickness $h>W_D$, the conduction through the sample is shunted by a 'dead-layer' near the top surface and the device remains to be conductive in the OFF state. The thickness of this 'dead-layer' is approximately $h - W_D$ and the residual OFF-state conductance in the depletion regime is

$$G_{OFF} \approx pe\mu_h W(h - W_D)/L , \qquad (1)$$

where $L$, $W$ and $h$ are the length, width and thickness of the sample and $\mu_h$ is the hole (or majority carrier) mobility. Based on these discussions and the relatively large thickness of SnS employed here (50-100 nm), we see that the poor gate tuning performance in the positive gate voltage regime is caused by the shunt effect from holes in the surface dead-layer.

Examining Eq.1, one sees that 2D semiconductor samples with heavier doping and larger thickness $h$ are expected to have larger residual OFF-state conductance in the depletion regime and a poor ON-OFF ratio. SnS is known to be p-type due to rich intrinsic defects.[22,23,28] Photoluminescence (PL) measurements were performed to gain more insights on the impurity states. Fig. 4c displays PL spectra from the 85 nm SnS sample at variable temperatures. A broad luminescence band centered at ~1.7 eV becomes better defined when temperature is lowered and is clearly seen at 200 K. As noted in prior PL study on SnS films,[35] this PL band is likely from the radiative defect states, attesting the existence of rich defect states in SnS studied here. Although fabricating SnS devices with thickness less than 50 nm turned out to be difficult, we performed a study to investigate the SnS FET device's performance upon successive thinning of the sample by dry etching. As displayed in Fig. 4d, after Ar ion etching at 300 W with Ar flow of 30 sccm at 0.1 Torr for 180 sec, the OFF-state conductance was greatly suppressed and a higher ON-OFF ratio (red curve vs. black curve in Fig. 4d) was obtained. Unfortunately, further ion etching over longer period of time likely created severe damage to the sample, resulting in sample's mobility being greatly reduced and the ON-OFF ratio cannot be further improved (green curve in Fig. 4d).

We found that compensating the acceptors with n-type doping on the surface of SnS flake is an effective way to suppress the conductance of surface dead layer and improve the ON-OFF ratio of device. Previously, Cs$_2$CO$_3$ was used to induce n-type doping in MoS$_2$ and black phosphorus.[36,37] We compare the gate modulated conductance of a SnS device before and after evaporating 5 nm Cs$_2$CO$_3$ on its surface. As demonstrated in Fig. 5a, at least an order of magnitude increase in the ON-OFF ratio was achieved in the same device after surface n-type doping with Cs$_2$CO$_3$. Hall effect measurements showed that Cs$_2$CO$_3$ doped sample has hole density about ten times lower than the pristine samples ($10^{10}$-4×$10^{11}$/cm$^2$ at room temperature in Fig. 5b vs. $10^{11}$-$10^{12}$/cm$^2$ in Fig. 3a). The sample also demonstrated high hole mobility (~100 cm$^2$/Vs at room $T$) which is comparable to the bulk value (Fig. 5c). It is also note-worthy that with the n-type Cs$_2$CO$_3$ doping compensating the p-type acceptors from intrinsic defects in SnS, not only the hole density was suppressed, there was also an increase in the trans-conductance and ON-state current (Fig. 5a). This is likely due to the reduced number of charged impurities (thus less carrier scattering and better mobility) after compensation. Note that the typical ON-OFF ratio of our SnS devices after n-type surface doping reached 100 but is significantly lower than that of TMD based (e.g. MoS$_2$) FETs. We believe that more elaborated control of doping and using thinner SnS samples will further improve the switching performance of SnS based 2D FET devices.

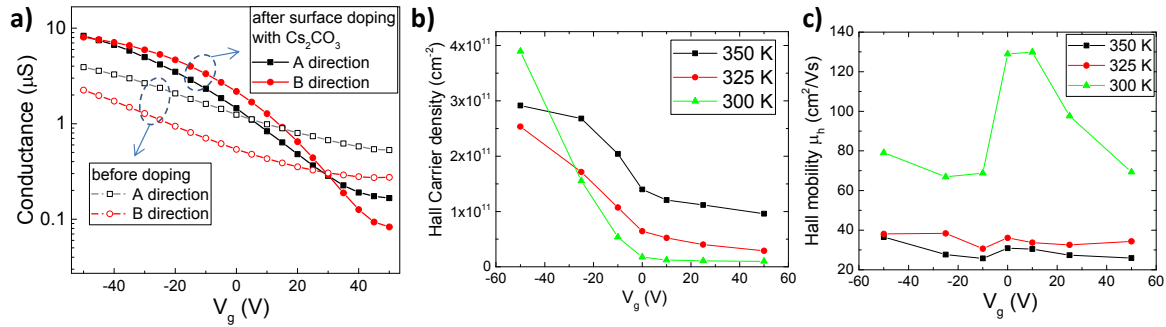

**Fig. 5.** a.) Comparison of 4-probe conductance *vs.* gate voltage at room temperature of SnS device between different direction and before and after doping with $Cs_2CO_3$. b.) Extracted Hall carrier density and d.) Hall mobility of the sample after doping with $Cs_2CO_3$.

## Experimental

### Bulk SnS single crystal growth

SnS single crystals were prepared from the Sn and S 5N (99.999%) purity compounds. The synthesis of the compounds was carried out in conical quartz ampoules evacuated to $10^{-4}$ Pa. The homogenization of the batches and synthesis of the compounds was carried out in a horizontal furnace at 900 °C for 48 h. The mixed crystals were grown by a vertical Bridgman method. Before pulling, the ampoules containing the melt were heat-treated at 900 °C for 24 h and when the melt filled the tip of ampoule, the ampoules were lowered through the temperature gradient at a rate of 0.1 mm/h. The amount of single crystalline areas was successfully increased by a reduction of the growth velocity from 5 mm/h to 0.5 mm/h. Approximately 2/3 of the whole sample volume is single crystalline, due to the still present stress during the phase transition. The obtained SnS single crystals were 3 cm long and 1.2 cm in diameter. They exhibited good cleavability in the direction perpendicular to the trigonal axis c, i.e. along the (001) plane. This preparation gave homogeneous single crystals with well-developed cleavage faces. These faces, being perpendicular to the c-axis, were always parallel with the direction of pulling, i.e. with the ampoule axis.

### Multilayer SnS Device Fabrication

The bulk SnS was then exfoliated onto degenerately doped Si substrates with silicon oxide or silicon nitride on surface. Electron beam lithography process with standard PMMA/MMA copolymer bilayer resist and metal deposition of Ni were subsequently used to create contacts for four-probe electrical transport characterization. Sample is a single crystal flake with atomically flat surface, and the sample generally has thickness ranging from 40-120 nm.

### Electrical transport characterization

Samples were measured in a Quantum Design PPMS (Physical Property Measurement System) or Lakeshore Cryogenic probe station. Hall measurements were conducted by varying magnetic field at various temperatures. For DC current-voltage characterization, a National Instrument PCI-6221 DAQ card integrated computer along with a BNC-2090 terminal block and a Stanford Research Systems SR-570 low noise current preamplifier were used to supply bias and measure current. A LabVIEW program was used to record data. For four-probe conductance and Hall resistance measurement, lockin technique was used where a 7 Hz sine wave voltage was applied to the sample. Current and four-probe voltage were measured by a SR-830 lockin amplifier.

### Raman and PL characterization

Raman and photoluminescence (PL) measurements were conducted using a Horiba Labram HR Raman microscope system. A 532 nm laser light was used. The laser was focused on sample surface using a 50× objective lens (spot diameter of about 2 μm). Instrument resolution was 0.5 $cm^{-1}$ for Raman measurements and 2 $cm^{-1}$ for PL measurements. Samples were mounted in an optical cryostat for variable temperature measurements.

## Conclusions

In conclusion, p-type multilayer SnS FET devices were fabricated and studied. Thermally activated behavior in the temperature dependent conductance and Hall carrier density demonstrated that the holes in devices originate from acceptors with ionization energy ~0.22 eV despite that no intentional doping was used in the SnS crystal. The devices showed good gate tunability in negative $V_g$ towards accumulation yet the gate tuning effect was weak in positive $V_g$ towards depletion. This contrast in the gate effect and the limited ON-OFF ratio could be attributed to the sample's thickness exceeding the maximum carrier screening length, allowing a residual conduction through a shunt layer near the top surface of sample. Through thinning the sample by etching or n-type surface doping, the device's ON-OFF ratio was improved by an order of magnitude. This work opens up routes towards high performance field effect devices based on IV-VI 2D mono-chalcogenides (SnS, SnSe *etc*) and points to the critical role of finite depletion depth in the switching performance (ON-OFF ratio, OFF-state conductance) of multilayer SnS devices. Since most 2D semiconductor FETs operate as depletion mode FET at room temperature, our findings here also offer insight to the understanding of other multilayer 2D semiconductor FETs. In particular, to obtain the best ON-OFF ratio, 2D semiconductor FETs should limit their thickness to within the Debye screening length (around 10 - 50 nm), beyond which there is a residual conductance from the top surface layer and a deteriorated ON-OFF ratio is expected. Similar conclusion regarding the deteriorated ON-OFF ratio with increasing thickness was reached in a recent study on multilayer $MoS_2$ FETs.[38]

## Acknowledgements

X.P.A.G. thanks the National Science Foundation (DMR-1151534) for its financial support and Jun Zhu for helpful discussions. Y.T.C. acknowledges the Ministry of Science and Technology of Taiwan (grant No. MOST-103-2627-M-002-009) for financial support. R.S. and F.C.C. acknowledge the support provided by the Academia Sinica research program on Nanoscience and Nanotechnology under project number NM004. Work at the University of Northern Iowa (UNI) is supported by the National Science Foundation (NSF, Grant No. DMR-1552482 (C.H.) and DMR-1410496 (R.H.)) and the ACS PRF grant (No. 53401-UNI10 (C.W.)). The low temperature equipment at UNI was acquired through the NSF MRI Grant (No. DMR-1337207). R. H. also acknowledges support by the UNI Faculty Summer Fellowship.

## References

1. K.S. Novoselov, A.K. Geim, S.V. Morozov, D. Jiang, M.I. Katsnelson, I.V. Grigorieva, S.V. Dubonos, A.A. Firsov. *Nature* 2005, **438**, 197–200.
2. Y. Zhang, Y.-W. Tan, H. L. Stormer, P. Kim, *Nature* 2005, **438**, 201–204.
3. Q.H. Wang, K. Kalantar-Zadeh, A. Kis, J.N. Coleman, M.S. Strano, *Nat. Nano.* 2012, **7**, 699–712.
4. A. K. Geim, I. V. Grigorieva, *Nature* 2013, **499**, 419–425.
5. M. Chhowalla, H.S. Shin, G. Eda, L.-J. Li, K.P. Loh, H. Zhang, *Nat. Chem.* 2013, **5**, 263–275.
6. X. Duan, C. Wang, A. Pan, R. Yu, X. Duan, *Chem. Soc. Rev.* 2015, **44**, 8859–8876.
7. K. F. Mak, J. Shan, *Nat. Photon.* 2016, **10**, 216–226.
8. B. Radisavljevic, A. Radenovic, J. Brivio, V. Giacometti, A. Kis, *Nat. Nano.* 2011, **6**, 147–150.
9. S. Ghatak, A. N. Pal, A. Ghosh, *ACS Nano* 2011, **5**, 7707–7712.
10. M. S. Fuhrer, J. Hone, *Nat. Nano.* 2013, **8**, 146–147.
11. S. Larentis, B. Fallahazad, E. Tutuc, *Appl. Phys. Lett.* 2012, **101**, 223104.
12. S.R. Das, J. Kwon, A. Prakash, C.J. Delker, S. Das, D.B. Janes, *Appl. Phys. Lett.* 2015, **106**, 83507.
13. W.S. Hwang, M. Remskar, R. Yan, V. Protasenko, K. Tahy, S.D. Chae, P. Zhao, A. Konar, H. (Grace) Xing, A. Seabaugh, D. Jena, *Appl. Phys. Lett.* 2012, **101**, 13107.
14. M.W. Iqbal, M.Z. Iqbal, M.F. Khan, M.A. Shehzad, Y. Seo, J.H. Park, C. Hwang, J. Eom, *Sci. Rep.* 2015, **5**, 10699.
15. N.R. Pradhan, D. Rhodes, S. Memaran, J.M. Poumirol, D. Smirnov, S. Talapatra, S. Feng, N. Perea-Lopez, A.L. Elias, M. Terrones, P.M. Ajayan, L. Balicas, *Sci. Rep.* 2015, **5**, 8979.
16. H. Liu, J. Chen, H. Yu, F. Yang, L. Jiao, G.-B. Liu, W. Ho, C. Gao, J. Jia, W. Yao, M. Xie, *Nat. Comm.* 2015, **6**, 8180.
17. L. Li, Y. Yu, G.J. Ye, Q. Ge, X. Ou, H. Wu, D. Feng, X.H. Chen, Y. Zhang, *Nat. Nano.* 2014, **9**, 372–377.
18. H. Liu, A. Neal, Z. Zhu, Z. Luo, X. Xu, D. Tomanek, and P. Ye, *ACS Nano*, 2014, **8** (4), 4033-40471.
19. S. Sucharitakul, N.J. Goble, U.R. Kumar, R. Sankar, Z.A. Bogorad, F.-C. Chou, Y.-T. Chen, X.P.A. Gao, *Nano Lett.* 2015, **15**, 3815–3819.
20. S. R. Tamalampudi, *et al.*, *Nano Lett.* 2014, **14**, 2800–2806.
21. H. Chuang, B. Chamlagain, M. Koehler, M. Perera, J. Yan, D. Madrus, D. Tomanek, and Z. Zhou, *Nano Lett.* 2016, **16**, 1896–1902.
22. F.-Y. Ran, Z. Xiao, H. Hiramatsu, H. Hosono, T. Kamiya, *Appl. Phys. Lett.* 2014, **104**, 72106.
23. F.-Y. Ran, Z. Xiao, Y. Toda, H. Hiramatsu, H. Hosono, T. Kamiya, *Sci. Rep.* 2015, **5**, 10428.
24. L.-D. Zhao, S.-H. Lo, Y. Zhang, H. Sun, G. Tan, C. Uher, C. Wolverton, V.P. Dravid, M.G. Kanatzidis, *Nature* 2014, **508**, 373–377.
25. G. Shi, E. Kioupakis, *J. of Appl. Phys.* 2015, **117**, 65103.
26. R. Guo, X. Wang, Y. Kuang, B. Huang, *Phys. Rev. B* 2015, **92**, 115202.
27. S. Zhao, H. Wang, Y. Zhou, L. Liao, Y. Jiang, X. Yang, G. Chen, M. Lin, Y. Wang, H. Peng, Z. Liu, *Nano Res.* 2015, **8**, 288–295.
28. W. Albers, C. Haas, H. J. Vink, and J. D. Wasscher, *J. Appl. Phys.* 1961, **32**, 2220.
29. A. S. Rodin, L. C. Gomes, A. Carvalho, A. H. Castro Neto, *Phys. Rev. B* 2016, **93**, 45431.
30. H. R. Chandrasekhar, R. G. Humphreys, U. Zwick, M. Cardona, *Phys. Rev. B* 1977, **15**, 2177–2183.
31. T. Raadik, M. Grossberg, J. Raudoja, R. Traksmaa, J. Krustok, *J. of Phys. and Chem. of Sol.* 2013, **74**, 1683–1685.
32. J. Wang, D. Rhodes, S. Feng, M.A.T. Nguyen, K. Watanabe, T. Taniguchi, T.E. Mallouk, M. Terrones, L. Balicas, J. Zhu, *Appl. Phys. Lett*. 2015, **106**, 152104.
33. S.A. Dayeh, D.P.R. Aplin, X. Zhou, P.K.L. Yu, E.T. Yu, D. Wang, *Small* 2007, **3**, 326–332.
34. S. M. Sze, K. Ng. Kwok, Physics of Semiconductor Devices, 2007, 3rd Edition, Wiley.
35. M. Devika, N. Koteeswara Reddy, M. Prashantha, K. Ramesh, S. Venkatramana Reddy, Y.B. Hahn, K.R. Gunasekhar, *Phys. Stat. Sol. (a)* 2010, **207**, 1864–1869.
36. J. Lin, C. Han, F. Wang, R. Wang, D. Xiang, S. Qin, X.-A. Zhang, L. Wang, H. Zhang, A. T.S. Wee, W. Chen, *ACS Nano*, 2014, **8** (5), 5323–5329.
37. D. Xiang, C. Han, J. Wu, S. Zhong, Y. Liu, J. Lin, X.-A. Zhang, W. P. Hu, B. Özyilmaz, A.H.C. Neto, A.T.S. Wee, W. Chen, *Nat. Comm.* 2015, **6**, 6485.
38. Y. Zhang, H. Li, H. Wang, H. Xie, R. Liu, S.-L. Zhang,; Z.-J. Qiu, *Sci. Rep.* 2016, **6**, 29615.

*Supplementary Information*

# Screening limited switching performance of multilayer 2D semiconductor FETs: the case for SnS

Sukrit Sucharitakul,[a] U. Rajesh Kumar,[bc] Raman Sankar,[de] Fang-Cheng Chou,[d] Yit-Tsong Chen,[bc] Chuhan Wang,[f] Cai He,[f] Rui He,[f] and Xuan P. A. Gao[*a]

| | |
|---|---|
| a. | *Department of Physics, Case Western Reserve University, Cleveland OH 44106, USA.* |
| b. | *Department of Chemistry, National Taiwan University, Taipei 10617, Taiwan.* |
| c. | *Institute of Atomic and Molecular Sciences, Academia Sinica, Taipei 10617, Taiwan.* |
| d. | *Center for Condensed Matter Sciences, National Taiwan University, Taipei 10617, Taiwan.* |
| e. | *Institute of Physics, Academia Sinica, Taipei 11529, Taiwan.* |
| f. | *Department of Physics, University of Northern Iowa, Cedar Falls, Iowa 50614, USA* |

*Email: xuan.gao@case.edu

**Anisotropic Hall mobility of SnS multilayer devices**

Anisotropic conductance was observed in SnS nanoflake van der Pauw devices, suggesting the holes have anisotropic transport mobility. The direction dependent Hall mobility for the 60 nm thick SnS device discussed in Figure 2 and 3 of the main manuscript is analyzed and displayed in Figure S1.

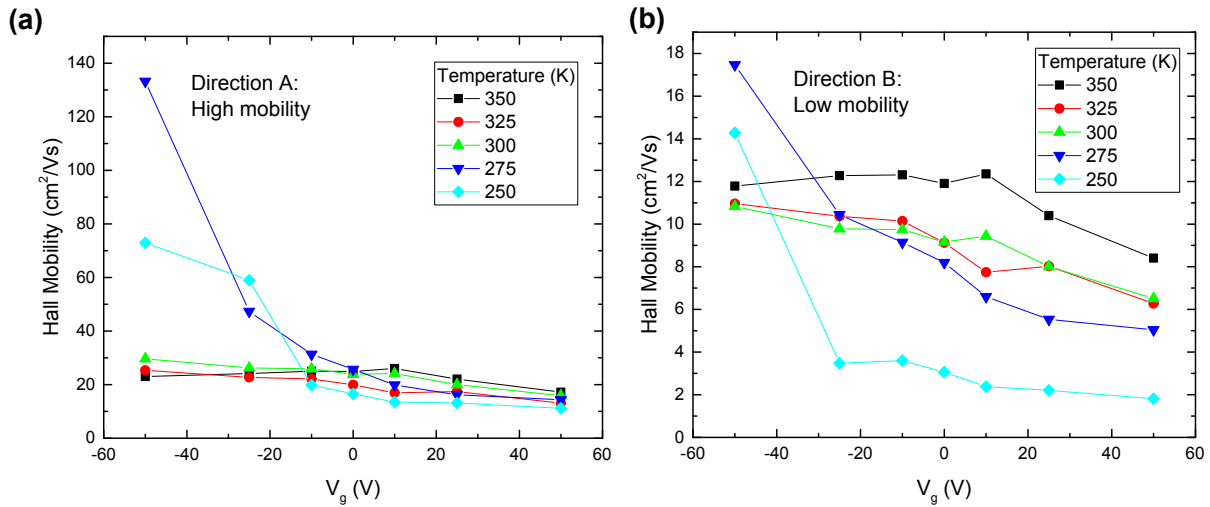

**Fig. S1**. Hall mobility for holes in a 60 nm thick SnS FET device along the high mobility or A-direction (a), and the low mobility or B-direction (b) *vs.* the backgate voltage at different temperatures.